\documentclass[lettersize,journal]{IEEEtran}
\usepackage{amsmath,amsfonts}
\usepackage{algorithmic}
\usepackage{algorithm}
\usepackage{array}
\usepackage[caption=false,font=normalsize,labelfont=sf,textfont=sf]{subfig}
\usepackage{textcomp}
\usepackage{stfloats}
\usepackage{url}
\usepackage{verbatim}
\usepackage{graphicx}
\usepackage{cite}
\usepackage{xcolor}
\usepackage{multirow}
\usepackage{soul}
\usepackage{threeparttable}

\usepackage{subfiles}
\usepackage[acronym]{glossaries}
\makeglossaries
\newacronym{ieee}{IEEE}{Institute of Electrical and Electronics Engineers}
\newacronym{sg}{SG}{Study Group}
\newacronym{wng}{WNG}{Wireless Next Generation}
\newacronym{sc}{SC}{Standing Committee}
\newacronym{tig}{TIG}{Topic Interest Group}
\newacronym{xr}{XR}{eXtended Reality}
\newacronym{he}{HE}{High Efficiency}
\newacronym{qam}{QAM}{Quadrature Amplitude Modulation}
\newacronym{llm}{LLM}{Large Language Model}
\newacronym{gpu}{GPU}{Graphics Processing Unit}
\newacronym{tpu}{TPU}{Tensor Processing Unit}
\newacronym{tops}{TOPS}{Tera Operations Per Second}
\newacronym{npu}{NPU}{Neural Processing Unit}
\newacronym{apu}{APU}{Accelerated Processing Unit}
\newacronym{par}{PAR}{Project Authorization Request}
\newacronym{win}{WIN}{WLAN Intelligent Networking}
\newacronym{vr}{VR}{Virtual Reality}
\newacronym{ar}{AR}{Augmented Reality}
\newacronym{obss}{OBSS}{Overlapping Basic Service Set}
\newacronym{iot}{IoT}{Internet of Things}
\newacronym[plural=APs]{ap}{AP}{Access Point}
\newacronym{csmaca}{CSMA/CA}{Carrier Sense Multiple Access with Collision Avoidance}
\newacronym{lbt}{LBT}{Listen-Before-Talk}
\newacronym[plural=BSSs]{bss}{BSS}{Basic Service Set}
\newacronym{mlo}{MLO}{Multi-Link Operation}
\newacronym{sr}{SR}{Spatial Reuse}
\newacronym{obsspd}{OBSS/PD}{Overlapping Basic Service Set/Packet Detect}
\newacronym[plural=STAs]{sta}{STA}{Station}
\newacronym{snr}{SNR}{Signal-to-Noise-Ratio}
\newacronym{tgbx}{TGbx}{Task Group bx}
\newacronym{tgbn}{TGbn}{Task Group bn}
\newacronym{tgbf}{TGbf}{Task Group bf}
\newacronym{tgbq}{TGbq}{Task Group bq}
\newacronym{tgbt}{TGbt}{Task Group bt}
\newacronym{tgbu}{TGbu}{Task Group bu}
\newacronym{aio}{AIO}{AI Offload}
\newacronym{aifsn}{AIFSN}{Arbitration Inter-Frame Space Number}
\newacronym{sdo}{SDO}{Standards Development Organization}
\newacronym{isac}{ISAC}{Integrated Sensing and Communication}
\newacronym{ldpc}{LDPC}{Low-Density Parity-Check}
\newacronym{sae}{SAE}{Simultaneous Authentication of Equals}
\newacronym{owe}{OWE}{Opportunistic Wireless Encryption}
\newacronym{fils}{FILS}{Fast Initial Link Setup}
\newacronym{pqc}{PQC}{Post-Quantum Cryptography}
\newacronym{wan}{WAN}{Wide Area Network}
\newacronym{mtp}{MTP}{Motion-to-Photon}
\newacronym{dl}{DL}{Deep Learning}
\newacronym{mcp}{MCP}{Model Context Protocol}
\newacronym{a2a}{A2A}{Agent2Agent Protocol}
\newacronym{itu}{ITU}{International Telecommunication Union}
\newacronym{ml-kem}{ML-KEM}{Module-Lattice Key Encapsulation Mechanism}
\newacronym{ml-dsa}{ML-DSA}{Module-Lattice Digital Signature Algorithm}
\newacronym{cosr}{Co-SR}{Coordinated Spatial Reuse}
\newacronym{cobf}{Co-BF}{Coordinated Beamforming}
\newacronym{wmm}{WMM}{Wi-Fi Multimedia}
\newacronym{marl}{MARL}{Multi-Agent Reinforcement Learning}
\newacronym{fl}{FL}{Federated Learning}
\newacronym{nn}{NN}{Neural Network}
\newacronym{ds}{DS}{Distribution System}
\newacronym{tflops}{TFLOPS}{Tera Floating-point Operations Per Second}
\newacronym{cortwt}{Co-RTWT}{Coordinated R-TWT}
\newacronym{cotdma}{Co-TDMA}{Coordinated Time Division Multiple Access}
\newacronym{jt}{JT}{Joint Transmissions}
\newacronym{coofdma}{Co-OFDMA}{Coordinated OFDMA}
\newacronym{mapc}{MAPC}{Multi-Access Point Coordination}
\newacronym{txop}{TXOP}{Transmission Opportunity}
\newacronym{ecsr}{ECSR}{Enhanced Coordinated Spatial Reuse}
\newacronym{rssi}{RSSI}{Received Signal Strength Indicator}
\newacronym{mcs}{MCS}{Modulation and Coding Scheme}
\newacronym{dcf}{DCF}{Distributed Coordination Function}
\newacronym{sinr}{SINR}{Signal-to-Interference-plus-Noise Ratio}
\newacronym{rrm}{RRM}{Radio Resource Management}
\newacronym{rts}{RTS}{Request-to-Send}
\newacronym{cts}{CTS}{Clear-to-Send}
\newacronym{arq}{ARQ}{Automatic Repeat Request}
\newacronym{ampdu}{A-MPDU}{Aggregate MAC Protocol Data Unit}
\newacronym{ai}{AI}{Artificial Intelligence}
\newacronym{ml}{ML}{Machine Learning}
\newacronym{phy}{PHY}{Physical}
\newacronym{mac}{MAC}{Medium Access Control}
\newacronym{uhr}{UHR}{Ultra-High Reliability}
\newacronym{sifs}{SIFS}{Short Interframe Space}
\newacronym{difs}{DIFS}{DCF Interframe Space}
\newacronym{aifs}{AIFS}{Arbitration Interframe Space}
\newacronym{wds}{WDS}{Wireless Distribution System}
\newacronym{ssid}{SSID}{Service Set Identifier}
\newacronym{ess}{ESS}{Extended Service Set}
\newacronym{mbss}{MBSS}{Mesh Basic Service Set}
\newacronym{wlc}{WLC}{Wireless LAN Controller}
\newacronym{wlan}{WLAN}{Wireless Local Area Network}
\newacronym{capwap}{CAPWAP}{Control And Provisioning of Wireless Access Points}
\newacronym{ietf}{IETF}{Internet Engineering Task Force}
\newacronym{rfc}{RFC}{Request for Comments}
\newacronym{isp}{ISP}{Internet Service Provider}
\newacronym{cpe}{CPE}{Customer-Premises Equipment}
\newacronym{usp}{USP}{User Services Platform}
\newacronym{mqtt}{MQTT}{Message Queuing Telemetry Transport}
\newacronym{bssid}{BSSID}{Basic Service Set Identifier}
\newacronym{tpc}{TPC}{Transmit Power Control}
\newacronym{cocr}{Co-CR}{Coordinated Channel Recommendation}
\newacronym{pasn}{PASN}{Pre-Association Security Negotiation}
\newacronym{ltf}{LTF}{Long Training Field}
\newacronym{gi}{GI}{Guard Interval}
\newacronym{twt}{TWT}{Target Wake Time}
\newacronym{sp}{SP}{Service Period}
\newacronym{rtwt}{R-TWT}{Restricted Target Wake Time}
\newacronym{aid}{AID}{Association ID}
\newacronym{pmksa}{PMKSA}{Pairwise Master Key Security Association}
\newacronym{ptksa}{PTKSA}{Pairwise Transient Key Security Association}
\newacronym{rsne}{RSNE}{Robust Security Network Element}
\newacronym{rsnxe}{RSNXE}{RSN eXtension Element}
\newacronym{mic}{MIC}{Message Integrity Check}
\newacronym{back}{BACK}{Block Acknowledgment}
\newacronym{ppdu}{PPDU}{Physical Layer Protocol Data Unit}
\newacronym{csi}{CSI}{Channel State Information}
\newacronym{ndp}{NDP}{Null Data Packet}
\newacronym{bsrp}{BSRP}{Buffer Status Report Poll}
\newacronym{ntb}{NTB}{Non-Trigger Based}
\newacronym{icf}{ICF}{Initial Control Frame}
\newacronym{icr}{ICR}{Initial Control Response}
\newacronym{dps}{DPS}{Dynamic Power Save}
\newacronym{emlsr}{EMLSR}{Enhanced Multi-Link Single Radio}
\newacronym{ofdm}{OFDM}{Orthogonal Frequency-Division Multiplexing}
\newacronym{eht}{EHT}{Extremely High Throughput}
\newacronym{ndpa}{NDPA}{NDP Announcement}
\newacronym{tb}{TB}{Trigger-Based}
\newacronym{bfrp}{BFRP}{Beamforming Report Poll}
\newacronym{ofdma}{OFDMA}{Orthogonal Frequency-Division Multiple Access}
\newacronym{mu}{MU}{Multi-User}
\newacronym{txs}{TXS}{Triggered TXOP Sharing}
\newacronym{mpdu}{MPDU}{MAC Protocol Data Unit}
\newacronym{tsf}{TSF}{Time Synchronization Function}
\newacronym{p2p}{P2P}{Peer-to-Peer}
\newacronym{txspg}{TXSPG}{TXOP Sharing for P2P Groups}
\newacronym{scs}{SCS}{Stream Classification Service}
\newacronym{qos}{QoS}{Quality of Service}
\newacronym{ru}{RU}{Resource Unit}
\newacronym{mimo}{MIMO}{Multiple Input Multiple Output}
\newacronym{cst}{CST}{Carrier Sense Threshold}
\newacronym{irm}{IRM}{Internal Regret Minimization}
\newacronym{ce}{CE}{Correlated Equilibrium}
\newacronym{ne}{NE}{Nash Equilibrium}
\newacronym{cca}{CCA}{Clear Channel Assessment}
\newacronym{psr}{PSR}{Parametrized Spatial Reuse}
\newacronym{mab}{MAB}{Multi-Armed Bandit}
\newacronym{ma}{MA}{Multi-Agent}
\newacronym{drl}{DRL}{Deep Reinforcement Learning}
\newacronym{rl}{RL}{Reinforcement Learning}
\newacronym{npca}{NPCA}{Non-Primary Channel Access}
\newacronym{dso}{DSO}{Dynamic Subband Operation}
\newacronym{edca}{EDCA}{Enhanced Distributed Channel Access}
\newacronym{dru}{DRU}{Distributed Resource Units}
\newacronym{fsm}{FSM}{Finite State Machine}
\newacronym{ack}{ACK}{acknowledgment}
\newacronym{nav}{NAV}{Network Allocation Vector}
\newacronym{tf}{TF}{Trigger Frame}
\newacronym{ula}{ULA}{Uniform Linear Array}
\newacronym{zf}{ZF}{Zero Forcing}
\newacronym{beb}{BEB}{Binary Exponential Backoff}
\newacronym{vo}{VO}{Voice}
\newacronym{vi}{VI}{Video}
\newacronym{be}{BE}{Best Effort}
\newacronym{bk}{BK}{Background}
\newacronym{ac}{AC}{Access Category}
\newacronym{cw}{CW}{Contention Window}
\newacronym{iyt}{IYT}{It's Your Turn}
\newacronym{eca}{ECA}{Enhanced Collision Avoidance}
\newacronym{dcb}{DCB}{Dynamic Channel Bonding}
\newacronym{dnn}{DNN}{Deep Neural Network}
\newacronym{gnn}{GNN}{Graph Neural Network}
\newacronym{psm}{PSM}{Power Save Mode}
\newacronym{dqn}{DQN}{Deep Q-Network}
\newacronym{ppo}{PPO}{Proximal Policy Optimization}
\newacronym{maddpg}{MADDPG}{Multi-Agent Deep Deterministic Policy Gradient}
\newacronym{murts}{MU-RTS}{Multi-User Request-to-Send}

\usepackage{booktabs}

\def\BibTeX{{\rm B\kern-.05em{\sc i\kern-.025em b}\kern-.08em
    T\kern-.1667em\lower.7ex\hbox{E}\kern-.125emX}}

\begin{document}

% This code is to reduce the list of authors by using et. al:
\bstctlcite{IEEEexample:BSTcontrol}

\title{IEEE 802.11bx -- WLAN Intelligent Networking (WIN):\\Toward an AI-Ready Wi-Fi 9}

\author{Francesc Wilhelmi, Katarzyna Kosek-Szott, Szymon Szott, and Boris Bellalta
        % <-this % stops a space
\thanks{F. Wilhelmi and B. Bellalta are with Universitat Pompeu Fabra (UPF), and K. Kosek-Szott and S. Szott are with AGH University of Krakow, Poland. This work was supported by the following projects: MLDR (Chist-ERA WAI 2022) PCI2023-145958-2 (MCIU/AEI/10.13039) and the National Science Centre, Poland (2023/05/Y/ST7/00004), TRUE Wi-Fi PID2024-155470NB-I00 (MICIU/AEI/10,13039/501100011033/FEDER,UE), ICREA Academia 2024 (00077 AGAUR), and MdM CEX2021-001195-M (MICIU/AEI/10.13039/501100011033).}% <-this % stops a space
}

\maketitle

\begin{abstract}
Wi-Fi~9 is expected to go beyond mere communication and provide new services such as sensing or computation. At this juncture, Artificial Intelligence (AI) is taking a leading role in the definition of the 802.11bx amendment, named WLAN Intelligent Networking (WIN). In this tutorial, we survey the recent progress made toward Wi-Fi~9 within IEEE 802.11 standardization, tracing the drivers and technological advances that motivate an AI-ready Wi-Fi~9. We then examine AI's role along three complementary dimensions, i.e., \textit{AI as a protocol} (AI is applied to Wi-Fi's PHY/MAC operation), \textit{AI as a platform} (Wi-Fi infrastructure is repurposed to provide AI computation), and \textit{AI as traffic} (AI flows call for new traffic-handling policies), and discuss candidate features and open challenges along each. As a concrete illustration of the \textit{AI as traffic} paradigm, we present a case study on AI traffic differentiation, where we explore a potential extension of the current Enhanced Distributed Channel Access (EDCA) to support new AI traffic flows.
\end{abstract}


\section{Introduction}
\label{sec:intro}

\glsunset{win}
\glsunset{wlan}

\IEEEPARstart{W}{LAN} Intelligent Networking (WIN) has been chosen to name the future standard for Wi-Fi~9 (IEEE 802.11bx), a true declaration of intent to finally consider \gls{ai} and \gls{ml} in Wi-Fi standardization. This is not a sudden pivot since the IEEE~802.11 \gls{ai}/\gls{ml} \gls{tig}, later elevated to a \gls{sc}, has spent several years studying the necessity and feasibility of making \glspl{wlan} intelligent~\cite{wilhelmi2024machine}. Simultaneously, advances in \gls{ai}, with \glspl{llm} at the forefront, are fueling new use cases and shifting how communications systems are used~\cite{ericsson2026uplink}, urging technologies like Wi-Fi to evolve accordingly. Furthermore, future applications like immersive \gls{xr} and integrated sensing, as well as new environments such as Industry 4.0 come with ever-increasing network requirements that have placed \gls{ai} and \gls{ml} at the center of the discussion.

Every generation of Wi-Fi has been motivated by a core idea that has driven the set of functionalities standardized by each one. For example, IEEE 802.11ax (Wi-Fi~6) targeted \textit{\gls{he}} by introducing \gls{ofdma} with trigger-based uplink multi-user transmissions, \gls{twt}, and \gls{sr}. IEEE 802.11be (Wi-Fi~7) focused on \textit{\gls{eht}} and introduced \gls{mlo} and 4K \gls{qam}. Currently, IEEE 802.11bn (Wi-Fi~8) is focusing on \textit{\gls{uhr}}~\cite{geraci2026wi}, with \gls{mapc} being the main innovation to achieve this goal~\cite{wilhelmi2026mapc}. Meanwhile, it is too early to provide specific performance targets for IEEE 802.11bx because its \gls{par} is in preparation and expected by 2027. However, the overall direction is already explicit in early \gls{wng}~\gls{sc} contributions and in the \gls{win} name itself. In this sense, \gls{ai} is established as Wi-Fi~9's core pillar rather than one feature among several, echoing the role reliability played for its predecessor.

In this tutorial, we discuss the potential directions to be adopted in 802.11bx standardization regarding \gls{ai} (either as an enabler or as a driver), which revolve around three plausible roles: \textit{i)} improving Wi-Fi operation (\textit{\gls{ai} as a protocol}), \textit{ii)} leveraging Wi-Fi infrastructure to run \gls{ai} (\textit{\gls{ai} as a platform}), and \textit{iii)} offering specialized network services and \gls{qos} management tailored to \gls{ai} workloads (\textit{\gls{ai} as traffic}). 

In the remainder of this tutorial, we first discuss the main drivers that are pushing next-generation Wi-Fi to be \gls{ai}-aware. Next, we describe the journey toward Wi-Fi~9 and review potential new features in standardization. Additionally, we present a case study for \gls{ai} traffic differentiation in Wi-Fi, supported by simulation results.  

\section{Drivers for an AI-Ready Wi-Fi 9}
\label{sec:drivers}

\begin{figure}[t!]
    \centering
    \includegraphics[width=\linewidth]{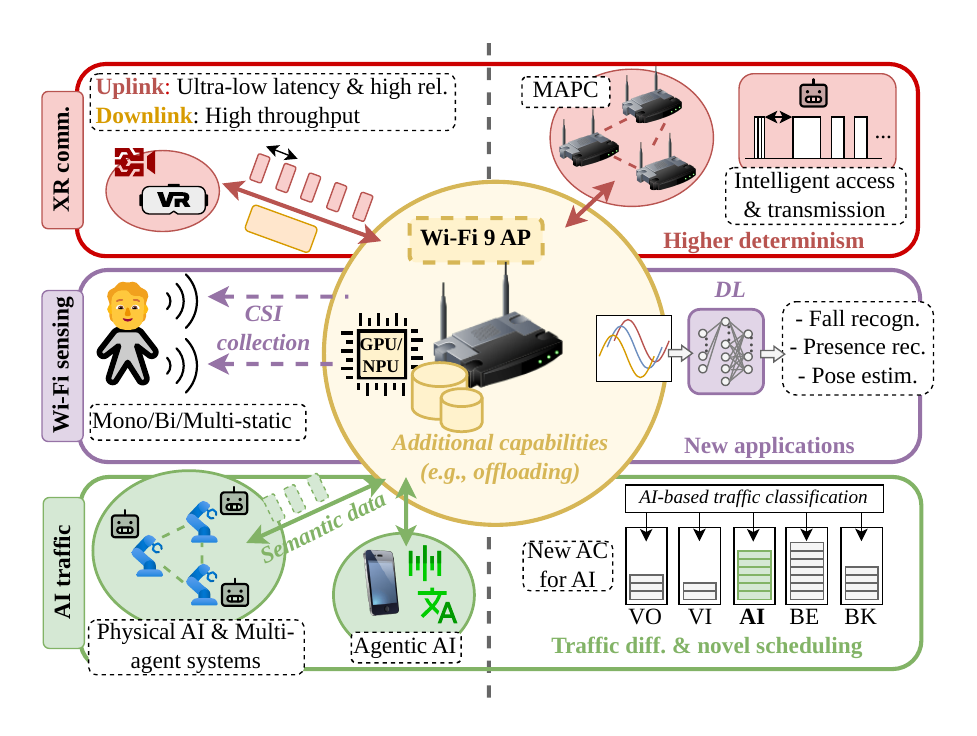}
    \caption{Drivers for an AI-ready Wi-Fi 9: (top) XR communications embody new, challenging networking paradigms, which require enhanced PHY/MAC schemes (e.g., MAPC, intelligent protocols); (middle) Wi-Fi sensing exemplifies the expanded capabilities of future Wi-Fi devices and potential interplay with AI and computation offloading; and (bottom) AI traffic calls for the redefinition of Wi-Fi traffic treatment policies, such as traffic differentiation.}
    \label{fig:drivers}
\end{figure}

In this section, we review the main drivers shaping the technology behind 802.11bx, categorized as application-, computation-, and traffic-centric drivers (Fig.~\ref{fig:drivers}).

\subsection{Application-Centric Drivers}

New applications have historically driven 802.11 innovation, since their requirements define what each generation must deliver. Next, we describe two prominent use cases for Wi-Fi that have been widely discussed in recent years, namely immersive communications and sensing.

\paragraph{Immersive Communications} 

\gls{xr} applications combine head-mounted displays and glasses, motion-tracking cameras, and haptic devices with stringent throughput, latency, and reliability constraints that, jointly, remain hard to guarantee by any existing Wi-Fi generation. In practice, split-rendered \gls{xr} streaming typically requires bitrates in the tens of Mb/s (e.g., $\sim$50\,Mb/s) and low, stable latency (e.g., $<$10-15\,ms)~\cite{carrascosa2025xr}. These extreme requirements are expected to increase in the future, with \gls{ai}-generated content for \gls{xr} (where \gls{ai} models create highly customized immersive experiences in real time) and continuous video models (where \gls{ai} analyzes real-time video streams to enhance the experience), which reinforces the need for novel features in Wi-Fi (e.g., \gls{ai}-driven channel access) and efficient offloading strategies.

\paragraph{Sensing} 

Wi-Fi \glspl{ap} will provide capabilities beyond communications, and a prominent domain is Wi-Fi sensing~\cite{meneghello2023toward}. By exploiting channel measurements in the form of \gls{csi}, Wi-Fi devices can detect presence, recognize gestures and activities, identify people, track motion, and monitor environments without requiring dedicated sensors or cameras. This capability unlocks a broad set of use cases in domains such as smart homes (e.g., fall monitoring for elderly care, pose estimation), enterprise and industry (e.g., intrusion detection), and healthcare (e.g., vital-sign and wellness monitoring). Many of these use cases are empowered by \gls{ai}, which excels at recognizing relevant patterns from complex \gls{csi} signals. For instance, deep learning models achieve 92--95\% accuracy when detecting human gestures from \gls{csi}~\cite{liu2026star}. Such results position sensing as an early candidate for running \gls{ai} inference on \glspl{ap}, since inference can potentially run alongside communication workloads on modest, power-constrained silicon.

\subsection{Computation-Centric Drivers}

A second class of drivers stems from where and how the computation associated with applications is actually performed. As \gls{ai} workloads increasingly demand resources beyond what individual devices can provide locally, the network itself becomes a candidate site for computation. Accordingly, we next describe three relevant computation-centric drivers, namely \textit{\gls{ai} computation offloading}, \textit{physical \gls{ai}}, and \textit{multi-agent systems}.

\paragraph{AI Computation Offloading}

According to Akamai, 50\% of \gls{ai} deployments in production struggle to maintain low latency at scale due to cloud infrastructure delays~\cite{akamai2026inference}. In response, new solutions increasingly rely on offloading inference and reasoning to the edge, rather than relying solely on distant cloud processing. However, this shift is only realizable if the edge devices have sufficient compute and memory. Although Wi-Fi \glspl{ap} have traditionally been designed for communication tasks, their proximity to users places them in a prime position to host computational workloads. This motivates the emergence of more computationally capable \glspl{ap} and the study of \gls{ai} offloading as an 802.11 standardized function (which we discuss later).

\paragraph{Physical AI and Multi-Agent Systems}

New \gls{ai} systems involving the physical world, such as autonomous robots and digital twins, depend on highly reliable, low-latency links, especially for cooperative tasks (e.g., requiring sub-millisecond round-trip delay between controllers and actuators~\cite{karamyshev2025towards}). When deployed as multi-agent systems, these physical entities rely on task decomposition to execute interdependent action sequences. Coordinating these parallel or sequential actions requires joint scheduling decisions to maximize the task success rate and minimize the overall completion time. This requirement pushes Wi-Fi networks to reason beyond pure data delivery, jointly accounting for task criticality, agent state, station capabilities, and available wireless resources. Consequently, future Wi-Fi \glspl{ap} will require schedulers that are aware not only of the traffic class but also of the structure of the tasks running on top.

\subsection{Traffic-Centric Drivers}

Finally, a third class of drivers refers to the nature of traffic from new applications. As \gls{ai}-generated and \gls{ai}-consumed data account for a growing share of network flows, their distinct statistical and semantic properties call into question assumptions that have long underpinned Wi-Fi's traffic handling. Therefore, we next describe \textit{\gls{ai} traffic} and \textit{semantic communications}.

\paragraph{AI Traffic}
New \gls{ai} applications are redefining how the Internet is used. First, agentic systems are rapidly becoming integral to our daily lives, serving as indispensable tools for entertainment and work. Many multi-modal commercial solutions (e.g., ChatGPT, Gemini, Claude) are widely adopted and increasingly shaping Internet traffic---by 2035, AI inference traffic could reach~25\% of total \gls{wan} traffic~\cite{cisco2026aiimpact}. Moreover, these solutions are changing the way end users use networks, given the intense human-agent and agent-agent interactions and their resulting network requirements. \gls{ai} flows are structurally different from web traffic, with longer sustained connections, smoother throughput, and more upstream data (uplink represents 9\% of \gls{ai} flows vs. 0.5\% for typical web traffic)~\cite{cisco2026aiimpact}. In addition, new trends in computation are shifting the workload to the edge, with more consumer-grade devices able to run powerful multi-agent models at user premises. This shift urges Wi-Fi networks, the dominant indoor access technology, to adapt accordingly.

\paragraph{Semantic Communications}
Communication systems were initially conceived to transmit bits as faithfully as possible, regardless of their meaning or purpose. Semantic communications aim to evolve the paradigm from classical bit-level capacity toward task-oriented effectiveness by focusing on the meaning and value of messages, which must be properly interpreted by a receiver to complete a given task~\cite{strinati2024goal}. Although not yet explicitly considered within Wi-Fi standardization, semantic communications might play an important role in future communications systems, as they alter how \gls{ai} agents use networks to interact with each other. New application-layer protocols such as \gls{mcp} and \gls{a2a}, which standardize how agents discover tools and peers, exchange context, and express intent, make the meaning of exchanged messages explicit. This opens the door to semantic-aware treatment of such traffic. Furthermore, the network could play a role in transmitting semantic information, in which case current all-or-nothing retransmission protocols in Wi-Fi might need to be revisited.

\section{The Road to WIN}
\label{sec:status}

\begin{figure*}
    \centering
    \includegraphics[width=.72\linewidth]{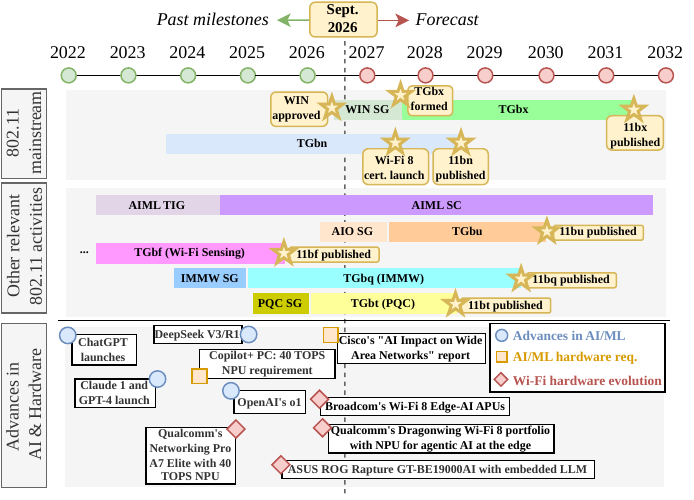}
    \caption{Timeline of WIN and related developments in IEEE 802.11, AI, and hardware.}
    \label{fig:timeline}
\end{figure*}

The path toward \gls{win} and \gls{tgbx} is the consolidation of several parallel developments, both within IEEE 802.11 and across the broader technology landscape. Figure~\ref{fig:timeline} describes the main events that motivated the development of 802.11bx and provides a forecasted roadmap of what is to come. 

\subsection{802.11 Mainstream}

Although 802.11bn is not yet finished (Draft~2.0 was approved in July 2026), the development of 802.11bx has begun in parallel. The incubator of next-generation Wi-Fi has been the \gls{wng}~\gls{sc}, whose activity in 2026 has spiked to host numerous discussions on new topics, including the role of \gls{ai}/\gls{ml}, novel backoff mechanisms, and open-source tools. The result is the \gls{win}~\gls{sg}, which now has the task of drafting the \gls{par} that will define Wi-Fi~9's formal scope. The definition of the \gls{par} (expected through 2027) will kick off the work of \gls{tgbx}, which will develop the 802.11bx specification, targeted for the early 2030s. Beyond any \gls{ai}-related features (whose inclusion remains uncertain), 802.11bx might follow up on 802.11bn's path and continue expanding features like \gls{mapc} (e.g., to support more than two coordinating \glspl{ap}) or propose new ones that have already been discussed (e.g., joint transmission, preemption)~\cite{karamyshev2025towards}. 

\subsection{Other Relevant 802.11 Activities}
\label{subsec:other}

The adoption of \gls{ai}/\gls{ml} in 802.11 traces back to the \gls{ai}/\gls{ml} \gls{tig} (approved in May 2022), which was initially tasked with identifying \gls{ai}/\gls{ml} use cases applicable to 802.11 systems and assessing their technical feasibility~\cite{wilhelmi2024machine}. The \gls{tig} report describing relevant use cases (\gls{csi} feedback compression, enhanced roaming, \gls{drl}-based channel access, \gls{ai}/\gls{ml}-driven \gls{mapc}, and \gls{ai}/\gls{ml} model sharing) signified the finalization of the \gls{tig}, which was then elevated to a permanent \gls{ai}/\gls{ml} \gls{sc}. Since then, the \gls{sc} has acted as the liaison point for \gls{ai}/\gls{ml} use cases, hosting contributions across 802.11 meetings.

A more recent milestone related to the adoption of \gls{ai}/\gls{ml} in Wi-Fi is the \gls{aio} \gls{sg} (approved in March 2026), which studied the offloading of compute-intensive \gls{ai} inference tasks to \glspl{ap} or edge devices. Progress has been unusually fast: the \gls{sg} completed its work and its \gls{par} led to the creation of \gls{tgbu} in July 2026. \gls{tgbu} will hold its inaugural meeting in January 2027. Based on the defined \gls{par}, \gls{tgbu} will work on mechanisms and interfaces that allow clients (e.g., \gls{vr} glasses) to offload computational tasks to the \gls{ap}. This could substantially decrease the latency of \gls{ai} and other types of applications compared to current cloud-based solutions. 

Beyond the work that is directly linked with \gls{ai}, other ongoing 802.11 activities are expected to work synergistically with \gls{ai} and contribute to the redefinition of Wi-Fi infrastructure. First, IEEE 802.11bf (developed by \gls{tgbf}), published in September 2025, was the first 802.11 amendment enabling devices to use Wi-Fi signals for sensing. Another relevant activity is \gls{tgbq} (Integrated mmWave), which aims to leverage \gls{mlo} to make use of high-frequency bands (42.5-48.4~GHz and 57-71~GHz) along with sub-7~GHz ones. While not directly related to \gls{ai}, the work from \gls{tgbq} will increase the complexity of Wi-Fi (e.g., beam-based operation, tighter \gls{ap} coordination) in return for higher performance, thereby expanding the capabilities of wireless infrastructure. Finally, \gls{tgbt} (\gls{pqc}) focuses on updating 802.11's authentication and key management suites against quantum-computing attacks, e.g., \gls{sae}, \gls{owe}, \gls{ap} PeerKey, \gls{fils}, and \gls{pasn}. Such a required level of security is expected to introduce profound changes to Wi-Fi infrastructure~\cite{hoque2026post}, creating a timely synergy between its significant data requirements and the emerging 802.11bu framework.

\subsection{Advances in AI and Hardware}

In recent years, we have witnessed fierce commercial competition in the field of \gls{ai} and chatbots. Since the appearance of ChatGPT in 2022, many \glspl{llm} have been released, pushing for higher performance (e.g., Claude 1 and GPT-4 in 2023), better reasoning capabilities (e.g., OpenAI's o1 in 2024), and higher efficiency (e.g., DeepSeek V3/R1 in 2024/2025). Driven by the requirements of \gls{ai} applications (e.g., Copilot+'s baseline of 40 \gls{tops} to run in real time), \gls{ai} computing has advanced rapidly. Consequently, more recently (from 2025 to today), there has been a shift in the development of hardware for computation, leading to chips that combine specialized processor types like \glspl{gpu} or \glspl{npu} with high-bandwidth memory in edge and consumer devices. This, together with the increase in \gls{ai} inference traffic~\cite{cisco2026aiimpact} and the proliferation of desktop-sized hardware that can run compact yet capable \gls{ai} models and \glspl{llm} (e.g., NVIDIA DGX Spark, Apple's Mac Studio), strengthens the case for next-generation Wi-Fi equipment and creates a unique market opportunity for Wi-Fi vendors and chip manufacturers. Some prominent examples, spanning from 2024 to today, are Qualcomm's Networking Pro A7 Elite, which has a 40-\gls{tops} \gls{npu}, Broadcom's Wi-Fi~8 chipsets (BCM49438 and BCM4918) with integrated neural engines, and ASUS's ROG Rapture GT-BE19000AI, which is capable of running \gls{ai} and \glspl{llm} locally.

\section{AI-driven Features for Wi-Fi 9}
\label{sec:features}

To address the emerging challenges and leverage the opportunities that \gls{ai} brings along, Wi-Fi will have to include novel features and functionalities. In this section, we describe the main directions that Wi-Fi may take for 802.11bx to become \gls{ai}-ready, differentiating \gls{ai}'s role as \textit{\gls{ai} as a protocol} (Wi-Fi adopts \gls{ai} to improve its PHY/MAC), \textit{\gls{ai} as a platform} (Wi-Fi infrastructure evolves to support \gls{ai} computations better), and \textit{\gls{ai} as traffic} (Wi-Fi treats new \gls{ai} traffic properly). Table~\ref{tab:features} summarizes these features along with their triggering drivers and the main related 802.11 groups.

\begin{table*}[t!]
\centering
\caption{Mapping between drivers, related standardization groups, AI's role, and potential new 802.11 features.}
\label{tab:features}
\resizebox{\textwidth}{!}{%
\begin{tabular}{@{}m{0.12\linewidth}m{0.18\linewidth}>{\centering\arraybackslash}m{0.09\linewidth}m{0.47\linewidth}@{}}
\toprule
\multicolumn{1}{c}{\textbf{Driver}} & \multicolumn{1}{c}{\textbf{Related 802.11 group(s)}} & \multicolumn{1}{c}{\textbf{AI's role}} & \multicolumn{1}{c}{\textbf{Potential new 802.11 feature(s)}} \\
\midrule
Immersive communications & TGbn, TGbx & Protocol & AI-driven MAPC beyond two APs to serve dense multi-user venues; AI-learned channel access (CW adaptation) for deterministic latency; AI-learned protocols; AI-based traffic forecasting\\
\addlinespace
Sensing & TGbf, AI/ML TIG/SC, AIO SG, TGbu & Platform & On-AP CSI-based sensing inference (early offloading use case given its modest compute footprint) \\
\addlinespace
\gls{ai} comp. offload & \gls{ai}/\gls{ml} TIG/SC, AIO SG, TGbu & Platform & AI/ML model sharing (MAC-layer signaling for model exchange); TGbu AI offload (discovery, compute advertisement, secure transport) \\
\addlinespace
Physical AI \& multi-agent sys. & AI/ML TIG/SC, WNG SC & Prot./Traffic & AI-driven MAPC for agent cooperation; task-dependent scheduling \\
\addlinespace
AI traffic & WNG SC & Traffic & New AI access category tuned to agentic traffic's statistical profile; task-aware scheduling of interdependent subtasks \\
\addlinespace
Semantic comm. & None & Traffic & Semantic-aware differentiation within a flow\textsuperscript{1} \\
\bottomrule
\end{tabular}
}
\raggedright\footnotesize\textsuperscript{1}Conceptual at this stage and not yet a concrete standardization proposal.
\end{table*}
 
\subsection{AI as a Protocol}
\label{sec:protocol}

Building on earlier progress in \gls{ai}/\gls{ml} \gls{tig}/\gls{sc}, \gls{ai} can be leveraged within Wi-Fi's PHY/MAC operation, either driving or assisting certain features for more performant and efficient operation. Based on the report delivered by the \gls{ai}/\gls{ml} \gls{tig}, some potential protocol-level use cases are \gls{csi} feedback compression for next-generation \glspl{wlan}, \gls{drl}-based channel access, enhanced roaming, and \gls{ai}-driven  \gls{mapc}~\cite{wilhelmi2024machine}. With the advent of Wi-Fi~8, some of these use cases can be expanded or reformulated, while new ones might be considered for Wi-Fi~9.

\paragraph{AI and MAPC}
The most direct extension from \gls{ai}/\gls{ml} \gls{tig}'s report is adopting \gls{ai} for \gls{mapc} beyond the two-\gls{ap} coordination introduced in 802.11bn, a potential way forward for 802.11bx. Both immersive communications and physical \gls{ai} drivers point to this underlying need, given that ever-increasing dense deployments are expected to bring multiple users (e.g., \gls{xr} glasses, robots) into the same venue, requiring multiple overlapping \glspl{ap} to serve them. As a result, coordinating beyond pairs of \glspl{ap} poses a critical challenge in terms of combinatorial complexity. In this regard, \gls{ai} can contribute in multiple ways: reducing the complexity of an exhaustive search (e.g., clustering, hierarchical learning), providing complementary information for decision-making (e.g., traffic forecasting), enabling collaboration while preserving privacy (e.g., \gls{fl}), or directly replacing static coordination rules with learned, topology-aware policies (e.g., \gls{gnn}, \gls{marl}). For the specific case of \gls{cobf}, \gls{ai}-based \gls{csi} compression becomes even more necessary when the number of coordinated \glspl{ap} is more than two. In that case, the overheads generated by channel estimation may become too heavy, eliminating the performance gains achieved with beamforming and null steering.

\paragraph{AI-Learned Protocols}
Another role that \gls{ai} can take is that of creator, rather than optimizer, participating in the autonomous design of communication and coexistence rules to use the spectrum more efficiently, and ultimately learning the structure of the mechanism itself (not only tuning its parameters). A timely example concerns channel access, whose core contention-based mechanism, i.e., the \gls{dcf}, has remained largely unchanged since the beginning of 802.11. Can \gls{cw} adaptation be driven by learned traffic patterns instead of static rules? Posing this question aims at architectural optimization while preserving \gls{dcf}'s fundamental status quo. A more ambitious goal would entrust \gls{ai} (e.g., under the \gls{rl} paradigm) to learn an outside-the-box protocol unconstrained by \gls{dcf}'s design, potentially replacing it entirely. This, combined with joint learning across PHY/MAC parameters (e.g., link adaptation, channel coding), could lead to fully \gls{ai}-learned protocols, a direction that aligns with the \gls{ai}-native air interface for 6G~\cite{hoydis2021toward}.

\paragraph{AI for Network Awareness}
\gls{ai} can also give Wi-Fi devices greater awareness of their radio environment and, in turn, enrich decision-making. Enhanced roaming, one of the \gls{ai}/\gls{ml} \gls{tig}'s use cases, exemplifies this. A client collects information from neighboring \glspl{ap} and feeds it to an \gls{ml} model that learns movement patterns and predicts the evolution of \gls{rssi}, producing a short and high-quality list of candidate \glspl{ap}~\cite{wilhelmi2024machine}. The same awareness principles apply to coexistence, a more recent direction being studied within the \gls{ai}/\gls{ml} \gls{sc}. As every 802.11 amendment adds capabilities on top of a backward-compatible frame structure, frame complexity and detection difficulty grow together, both across 802.11 generations and against unlicensed technologies sharing the same spectrum. A potential solution is deep-learning-based frame format detection, allowing a device to classify the kind of transmission it is observing before fully decoding it and adapt its behavior accordingly (e.g., differentiated \gls{sr} policies for legacy vs. non-802.11 frames).

\subsection{AI as a Platform}
\label{sec:platform}

A second direction envisions Wi-Fi infrastructure as a computation platform for \gls{ai}. Before the start of \gls{ai} Offload \gls{sg} and \gls{tgbu}, computation aspects had already been considered in 802.11 discussions. In particular, among the \gls{ai}/\gls{ml} \gls{tig}'s original use cases, \textit{\gls{ai}/\gls{ml} model sharing} was studied as an enabler of \gls{ai} over Wi-Fi, rather than as an \gls{ai}-driven optimization feature~\cite{wilhelmi2024machine}. Originally, the \gls{ai}/\gls{ml} model sharing use case was motivated by the large amount of information exchanged by emerging \gls{ai} (e.g., creating a training corpus), which was seen as a threat to the \gls{qos} of other traditional applications (e.g., voice, video). To address this, the \gls{tig} concluded that 802.11 should account for \gls{ml} model exchange overhead for better airtime distribution, potentially impacting the standard through architectural changes (e.g., to allow the exchange of \gls{ml} models at the MAC layer) and new signaling (e.g., to indicate support for \gls{ai}/\gls{ml} model sharing).

This subject has since gained new momentum and is being advanced through a dedicated standardization vehicle, \gls{tgbu}. In its \gls{par}, TGbu defines the scope of the project as providing the necessary MAC-layer transport and signaling support to enable secure \gls{ai} offloading services (e.g., sensing inference, generative \gls{xr} content, agentic reasoning) from a compute-capable device, including \glspl{ap}, clients, and any other device reachable through a \gls{ds}~\cite{ieee80211bu2026par}. Specifically, this involves the definition of new discovery capabilities for Wi-Fi devices, which must be able to advertise the offloading services they offer and their location, along with mechanisms to request offloading services and retrieve their results (including mutual trust enforcement). A preliminary framework from Qualcomm (presented as contribution ``802.11-26/1391'' in an \gls{ai} Offload \gls{sg} meeting) envisions a case where \gls{xr} glasses offload an  \gls{ai} workload to an \gls{ap}, which executes a model retrieved from a model repository. To realize this use case, the pertinent computation and communication requirements must be negotiated and agreed upon during the session setup.

While the direction is clear, several open questions remain for \gls{tgbu}. First, compute capability must be quantified in a standardized manner, which is challenging because compute capability is inherently heterogeneous (e.g., \gls{gpu} vs. \gls{npu}) and depends dynamically on the available resources (including memory status) rather than being a fixed hardware property. Second, \gls{ai} computation offloading adds another dimension for admission control and fairness, given that compute will represent another limited resource that, together with airtime, will need to be carefully allocated. Therefore, \gls{tgbu} will need to simultaneously define how \glspl{ap} accept/reject, queue, or renegotiate requests under contention from multiple stations, a problem with no direct precedent in 802.11 for compute resources. Third, since the \gls{par} explicitly refers to \emph{secure} offloading, extensive work is required on defining and incorporating cryptographic primitives and mutual authentication procedures, a task that synergizes well with \gls{tgbt} activities.

\subsection{AI as Traffic}
\label{sec:pipe}

Wi-Fi's channel access was engineered around web, file exchange, and multimedia traffic flows, which are characterized by being downlink-dominated, bursty, and often independent of each other. Agentic \gls{ai} traffic breaks these assumptions, bringing uplink-heavy interactions, sustained flows, and task-dependent exchanges, where multiple subtasks stem from a single logical request and travel through different agents before a result is provided. Scheduling dependent subtasks properly becomes a primary concern, as it can reduce the end-to-end latency by 20\% and improve the overall task reward by 80\% compared to task-unaware approaches~\cite{han2026task}. This motivates two complementary directions for \gls{ai} traffic treatment. One consists of introducing a new \gls{ac} for agentic \gls{ai} traffic, rather than approximating it within existing classes like voice or video (which instead sustain continuous, jitter-bounded streams) or best-effort (which is fully elastic and delay-tolerant). In this regard, agentic traffic's particularities (where an agent typically waits for a low-latency response) motivate treating \gls{ai} differently, as later explored. % in Section~\ref{sec:use_case}. 
The second direction is more fine-grained and relates to the physical \gls{ai} and multi-agent systems drivers. It consists of differentiating not just by \gls{ac}, but also by the dependency structure within a single \gls{ai} flow, so that subtasks on which others depend receive different treatment from non-critical ones.

\section{Case Study: AI Traffic Differentiation in Wi-Fi}
\label{sec:use_case}

To ground the \textit{\gls{ai} as traffic} direction described in Section~\ref{sec:pipe}, we study what a new candidate \gls{edca} \gls{ac} for \gls{ai} (AC\_AI) would need to properly support \gls{ai} traffic without degrading traditional \gls{vo}, \gls{vi}, and \gls{be} traffic. Simulations were performed using \texttt{Kom8ndor}.\footnote{\url{https://github.com/wn-upf/Komondor/tree/dev} (commit: 4e1cca5).}

\subsection{Scenario}

We evaluate \gls{ai} traffic differentiation in a \gls{bss} with multiple associated stations ($N_\text{STA} = 23$), each at $d_{\text{AP-STA}} = 3$\,m from the \gls{ap}. Three of the stations carry voice (AC\_VO), video (AC\_VI), and best-effort (AC\_BE) traffic, respectively, representing a high-contention baseline. The rest carry \gls{ai} traffic and represent a classroom of $N=20$ students using \glspl{llm}, corresponding to a $\sim$6\,Mb/s load distributed between uplink (prompts) and downlink (responses). Each session models an uplink, bursty \gls{llm} token-generation stream of small frames (4\,kbit) with inter-arrival times uniformly drawn from $[10,33]$\,ms during bursts of duration $[10,60]$\,s. Idle periods between bursts are also uniformly drawn from $[5,30]$\,s. The downlink follows the same model but divides the intra-burst rate and idle gaps by $N$ to reflect the aggregation of responses into a single stream.
Table~\ref{tab:sim_params} provides the main simulation parameters.

\begin{table}[t]
\centering
\caption{Simulation parameters.}
\label{tab:sim_params}
\resizebox{\columnwidth}{!}{%
\begin{tabular}{clc}
\toprule
& \textbf{Parameter} & \textbf{Value} \\
\midrule
\multirow{5}{*}{\rotatebox{90}{\textit{Scenario}}}
& Number of stations, $N_{\text{STA}}$ & 23 \\
& AP-STA distance, $d_{\text{AP-STA}}$ & 3~m \\
& VO/VI/BE traffic load, $\rho_\text{legacy}$ & 5/40/80 Mb/s (Poisson) \\
& VO/VI/BE/AI data frame length, $L$ & 1.2/12/12/4 kbit \\
& Simulation time, $T$ & 300~s \\
\midrule
\multirow{6}{*}{\rotatebox{90}{\textit{PHY/MAC}}}
& Central frequency, $f_c$ & 5~GHz \\
& Bandwidth, $B$ & 80~MHz \\
& Default transmit power, $P$ & 20~dBm \\
& Pathloss model, $PL$ & TGax residential (see~\cite{wilhelmi2026mapc}) \\
& Carrier sense threshold, $S$ & $-82$ dBm \\
& Maximum aggregated MPDUs, $N_{\text{agg}}$ & 64 \\
\midrule
\multirow{4}{*}{\rotatebox{90}{\textit{EDCA}\textsuperscript{1}}}
& Min. contention window, $CW_{\min}$      & \{4, 16, 8, 32\} \\
& Max. contention window, $CW_{\max}$      & \{16, 32, 64, 1024\} \\
& Arbitration inter-frame space number, AIFSN   & \{2, 2, 4, 5\} \\
& TXOP duration limit\textsuperscript{2}, TXOP$_{\max}$ & \{1.504, 3.008, 0, 0\} ms \\
\bottomrule
\end{tabular}
}
\raggedright\footnotesize\textsuperscript{1}Values listed in order AC\_VO / AC\_VI / \textbf{AC\_AI} / AC\_BE.\\
\raggedright\footnotesize\textsuperscript{2}A TXOP limit of ``0'' denotes a single PPDU (can be A-MPDU) per access.
\end{table}

\subsection{Simulation Results}

The \gls{ac} parameters used throughout this evaluation are chosen so that AC\_AI's priority sits between AC\_VI and AC\_BE. Therefore, its AIFSN (4) is set between AC\_VI's (2) and AC\_BE's (5) to preserve the hierarchy AC\_VO $\succ$ AC\_VI $\succ$ AC\_AI $\succ$ AC\_BE $\succ$ AC\_BK, and its \gls{txop} limit is set to 0 to let AC\_AI empty its buffered small \gls{ai} traffic payloads in one transmission using \gls{ampdu} aggregation. For the \gls{cw}, we use $CW_{\min}=8$ to give AC\_AI faster access under light contention and $CW_{\max}=64$ to limit its aggressiveness as contention grows and improve delivery reliability.

\begin{figure*}[t!]
    \centering
    \includegraphics[width=.9\linewidth]{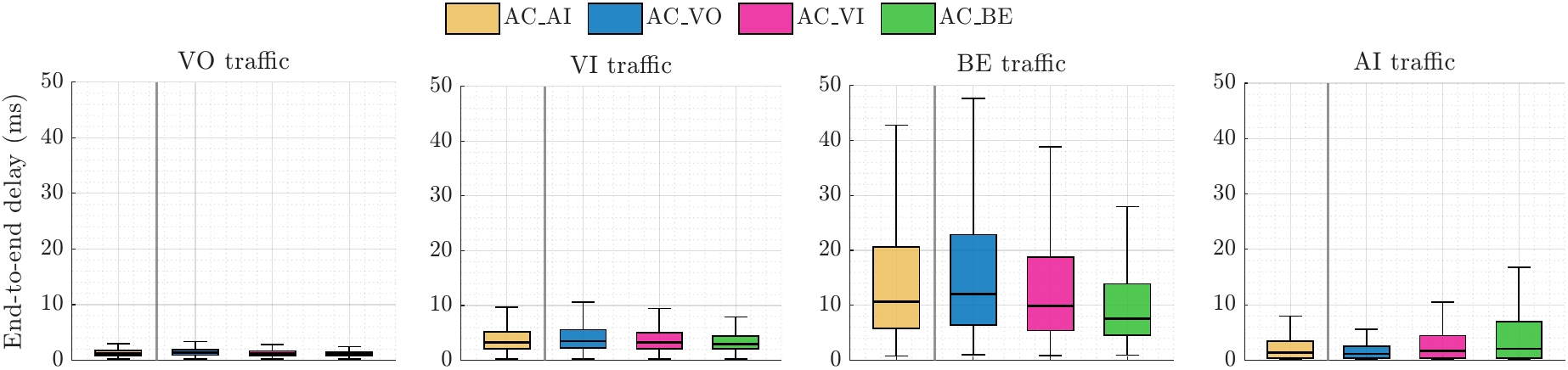}
    \caption{Distribution of the end-to-end delay (in ms) experienced by each traffic flow across the different AC assignments to AI traffic. For each box, the lower, central, and upper lines indicate the P25, P50, and P75, respectively, while the whiskers show the values at 1.5$\times$IQR of the nearest quartile.}
    \label{fig:plot_1_access_category}
\end{figure*}

Figure~\ref{fig:plot_1_access_category} shows the distribution of the end-to-end delay (measured from packet generation to successful delivery) for each traffic flow (\gls{vo}, \gls{vi}, \gls{be}, and \gls{ai}) when \gls{ai} packets are assigned to each of the four considered \glspl{ac}. The results show that no single assignment is safe for every traffic type, motivating the introduction of a new \gls{ac} for \gls{ai} rather than reusing an existing one. In particular, AC\_AI avoids two harmful assignments. First, mapping \gls{ai} traffic to AC\_VO minimizes \gls{ai}'s own delay (1.15\,ms) but inflicts the worst outcome on \gls{be}, whose median rises to 12.04\,ms and whose tail latency (upper whisker) reaches 47.59\,ms, the highest of any \gls{ac} combination. Second, mapping it to AC\_BE protects \gls{be} (7.60\,ms median, 27.93\,ms tail) but increases \gls{ai}'s own median delay by 45\% (2.08\,ms) and more than doubles its tail (16.75\,ms) relative to AC\_AI (1.43\,ms and 7.99\,ms, respectively). Relative to AC\_VI, AC\_AI gives \gls{ai} a lower median delay (1.43\,ms vs.\ 1.71\,ms) and a lower tail (7.99\,ms vs.\ 10.49\,ms), but at the cost of higher median delay for \gls{vo}, \gls{vi}, and \gls{be} (+1\%, +2\%, and +8\%, respectively). Fortunately, \gls{edca} prioritization preserves \gls{qos} for the two highest priority classes (AC\_VO and AC\_VI), so that the impact of the new AC\_AI is minimal.

\section{Conclusions}
\label{sec:conclusions}

This tutorial examined the road toward Wi-Fi~9 from the perspective of \gls{ai}, an important principle behind \gls{win}. We reviewed the application-, computation-, and traffic-centric drivers pushing 802.11bx toward \gls{ai}-awareness, described relevant standardization activities and hardware trends converging on this direction, and discussed candidate features along three complementary roles \gls{ai} can take in future Wi-Fi, namely \textit{\gls{ai} as a protocol}, \textit{\gls{ai} as a platform}, and \textit{\gls{ai} as traffic}. To ground the \textit{\gls{ai} as traffic} dimension, we presented a case study on introducing a dedicated access category, which underscored the need for \gls{ai} traffic differentiation beyond current \gls{edca} categories. As the \gls{win} \gls{sg} moves toward drafting the 802.11bx \gls{par}, several questions remain open for an \gls{ai}-ready Wi-Fi~9, including how to reconcile \gls{ai}-driven and \gls{ai}-learned protocols with Wi-Fi's legacy and backward-compatible design, how to standardize and fairly allocate compute as a new network resource, and how to treat \gls{ai} traffic without penalizing legacy traffic classes. Answering them will decide whether Wi-Fi~9 will merely accommodate \gls{ai} traffic or become the first \gls{wlan} generation genuinely co-designed with the intelligent systems it supports.

%%%%%%%%%%%%%%%%%%%%%%%%%
%%%  BIBLIOGRAPHY    
%%%%%%%%%%%%%%%%%%%%%%%%%

\bibliographystyle{IEEEtran}
\bibliography{references}

\vspace{-1cm}

\begin{IEEEbiographynophoto}{Francesc Wilhelmi}
(Senior Member, IEEE) is currently a tenure-track professor at Universitat Pompeu Fabra (UPF). His main research interests are Wi-Fi technologies and their evolution, network simulators and network digital twinning, machine learning, decentralized learning, and distributed ledger technologies. In 2024, he received the IEEE ComSoc Best Young Professional Award.    
\end{IEEEbiographynophoto}

\vspace{-1cm}

\begin{IEEEbiographynophoto}{Katarzyna Kosek-Szott} 
is a Full Professor at AGH University of Krakow, Poland. Her research interests are in the area of wireless networks, with emphasis on novel 802.11 mechanisms, performance improvement with machine learning, and the coexistence of wireless technologies in unlicensed bands.
\end{IEEEbiographynophoto}

\vspace{-1cm}

\begin{IEEEbiographynophoto}{Szymon Szott} 
is a Full Professor at AGH University of Krakow, Poland. His professional interests are related to wireless local area networks (channel access, quality of service, security, inter-technology coexistence). He is a member of the IEEE 802.11 Working Group.
\end{IEEEbiographynophoto}

\vspace{-1cm}

\begin{IEEEbiographynophoto}{Boris Bellalta} (Senior Member, IEEE)
is a Professor at Universitat Pompeu Fabra (UPF), where he leads the Wireless Networking group. His research interests lie in the area of wireless networks and performance evaluation, with a particular emphasis on Wi-Fi technologies and machine learning-based adaptive systems.  
\end{IEEEbiographynophoto}

\vspace{-1cm}

\end{document}